\documentclass[aps,final,notitlepage,oneside,twocolumn,nobibnotes,nofootinbib,
superscriptaddress,noshowpacs,centertags]{revtex4-1}

\usepackage[utf8]{inputenc}
\usepackage[english]{babel}
\usepackage{graphicx}
\usepackage{latexsym}
\usepackage{amssymb}
\usepackage{amsmath}
\usepackage{float}

\begin{document}

\title{Mergers of Binary Primordial Black Holes in Evolving Dark Matter Halos}
\author{V. D. Stasenko}\thanks{e-mail: vdstasenko@mephi.ru}
\affiliation{National Research Nuclear University, MEPhI, Moscow, 115409 Russia}
\affiliation{Institute for Nuclear Research, Russian Academy of Sciences, Moscow, 117312 Russia}
\author{Yu. N. Eroshenko}\thanks{e-mail: eroshenko@inr.ac.ru}
\affiliation{Institute for Nuclear Research, Russian Academy of Sciences, Moscow, 117312 Russia}

\date{\today}

\begin{abstract}
The influence of a dark matter halo around pair of primordial black holes on their orbit evolution and the black hole merger rate is considered. Because of the nonspherical (nonradial) contraction of DM shells, each shell upon the first contraction passes through the halo center in the direction of the radius vector corresponding to zero angular momentum. Since the shell contraction is a continuous process, at each instant of time there is a nonzero dark matter density at the halo center. This density is determined by the influence of the tidal gravitational forces from inflationary density perturbations and from other primordial black holes. The scattering of dark matter particles by a pair of black holes leads to a loss of the energy of its orbital motion and to an accelerated pair merger. In the case of primordial black holes with masses $\sim30M_\odot$, the black hole merger rate in the presence of a dark matter halo is several times higher than that without such a halo.
\end{abstract}

\maketitle


\section{INTRODUCTION}

The idea about the formation of primordial black holes (PBHs) in the early Universe was proposed almost 60 years ago by~\cite{ZelNov67, Haw71, KhlPol0}, and several models of their formation have been proposed
since then~\cite{KhlPol0,BerKuzTka83,Jed97,DolSil93,RubKhlSak00,RubSakKhl01}. A multitude of studies of the possible role of PBHs in various astrophysical
processes and in cosmology have also been performed. As examples we will point out that PBHs can be seeds for the early formation of supermassive black holes in quasars, while PBHs with masses $M_{\rm PBH}\sim30M_\odot$ can explain some of the gravitational wave events observed by the LIGO/Virgo/KAGRA detectors~\cite{Naketal97,Ioketal98,Sasetal16,Doletal20}.

The ranges of masses $M_{\rm PBH}\sim10^{20}-10^{24}$~g and $M_{\rm PBH}\sim 10-10^3M_\odot$ in which PBHs can account for the dominant part of dark matter (DM) still remain, but at other masses PBHs can make only a small contribution $f\ll 1$ to the total DM density because of various constraints (for a review, see~\cite{CarKuh20}). In this paper we consider a model in which PBHs are responsible for a significant fraction of LIGO/Virgo/KAGRA events, accounting for a fraction $f\ll 1$ in DM, and will consider the influence of the remaining DM on the evolution of PBH pairs before their merger with the generation of gravitational wave bursts.

The interaction of DM with PBH pairs was considered in various aspects in a number of papers. The work~\cite{Ero16-1} pointed out that the cosmological DM density perturbations that arose at the inflation stage could influence the angular momentum of a PBH pair, slightly changing the final merger rate. DM spikes at the radiation-dominated stage~\cite{Ero16-2} and more massive DM halos at the dust-like stage~\cite{DokEro01,DokEro03} can arise around PBHs. In particular, such halos must have been formed not only around individual PBHs, but also around a PBH pair as a whole~\cite{Hayetal09}. The works~\cite{StaBel23,Sta24} considered the role of PBHs in the formation of density perturbations giving rise to large DM halos and the backreaction effect of such halos on PBH clustering.

The angular momentum of a PBH pair is determined primarily by the gravitational influence of the third nearest PBH. The work~\cite{Hayetal09} pointed out that this third PBH must exert a gravitational effect on the contraction of DM shells in the halo around the PBH pair, causing DM particles to fall into the loss cone of the PBH pair. This, in turn, leads to an additional loss of the energy of its orbital motion. The work~\cite{Hayetal09} made an order-of-magnitude estimate of this effect. In this paper we consider the influence of the third PBH on the DM halo in more detail and by taking into account the nonspherical halo contraction. We also consider the influence of inflationary DM density perturbations with scales comparable to the size of the halo itself on the evolution of the DM halo.

A nonzero DM density at the halo center because of the nonspherical contraction of DM shells with the passage of these shells through the halo center, where the PBH pair is located, is a new effect that has not been explicitly taken into account previously and that we consider within an approximate analytical model. The evolution of a DM halo by taking into account the particle angular momentum was considered in~\cite{SikTkaWan97}, where a self-similar solution was found for some special cases. The work~\cite{SikTkaWan97} pointed out that, according to the fixed point theorem, directions in which the angular momentum of DM particles is zero must exist in spherical DM shells. In these directions the particles from a DM shell pass through the halo center upon the first contraction. After several oscillations, the DM shell is smeared out and passes into a general spherically symmetric DM distribution. In other directions the angular momentum is nonzero, leading to the passage of particles by the center of the forming halo at some characteristic distance that should be found to calculate the density of the DM shell.

In contrast to~\cite{SikTkaWan97}, we do not average the angular momentum of the DM particles over the spherical shell, but consider the shell contraction in two directions: in the specified direction of zero angular momentum and in the direction of the root-mean-square (rms) angular momentum. The contraction in the first of these directions leads to the passage of the DM shell through the center of the DM halo in which the PBH pair is located, while the contraction in the second direction stops at some typical radius. As a result, the DM shell as it passes through the halo center may be approximately represented as a disk, as can be seen from Fig.~2 in~\cite{SikTkaWan97}. We calculate the characteristic sizes of this disk and its density.

Since the passage of DM shells through the center is a continuous process (only the direction of the passage velocity and the disk radius change with
time), the PBH pair at each instant is immersed in a DM flow with a density slowly changing with time, and we calculate the loss of the energy of its orbital motion through its interaction with the DM flow. Concurrently, we consider the influence of the third nearest PBH on the contracting DM shell. Note that~\cite{Hayetal09} ignored the fact that a DM halo also grows around the third PBH, whose mass should be summed with the mass of the third PBH.

The above processes of the passage of DM shells through the center of the forming halo and their influence on the pair could be fully taken into account in high-resolution numerical simulations, and calculations within numerical simulations were carried out by~\cite{TkaPilYep20,PilTkaIva22}. Although the numerical simulations include the passages of DM through the halo center, only the passage of sufficiently large DM clumps can be taken into account in the numerical simulations because of their low resolution, but the effect of the passage of a large amount of background DM in such simulations being considered by us is largely inaccessible to consideration. We will show that including the background DM increases the overall result (the PBH pair merger rate) by several more times compared to the simulation result of~\cite{PilTkaIva22}.

\section{THE DENSITY AT THE CENTER OF THE HALO SURROUNDING A PBH PAIR}

A region with a density $\sim\rho_{\rm eq}$ and a radius $\sim(3M_{\rm PBH}/(4\pi\rho_{\rm eq}))^{1/3}$ is formed around a PBH pair near the transition time $t_{\rm eq}$ of the Universe to the dust-like stage of evolution (hereafter we mark the quantities at the time $t_{\rm eq}$ by the subscript “eq” and the quantities at the present time $t_0$ by the subscript “0”). The DM particles from this region passing through the center will influence the evolution of the pair orbit for some time, but they will be ejected to wider orbits. Nevertheless, newer and newer DM shells will constantly flow onto the PBH pair, and the DM density near the center will always be nonzero. Let us find this density.

The equations of radial motion for the DM shells have a solution in the parametric form
\begin{equation}
r=r_s\cos^2\theta, \quad
\theta+\frac{1}{2}\sin2\theta=\frac{2}{3}\left(\frac{5\delta_{\text{eq}}}{3}
\right)^{3/2}\frac{t-t_s}{t_{\text{eq}}}, 
\label{param}
\end{equation}
where the stopping radius of the shell at the time of its maximum expansion $t_s$ is denoted by $r_s$,
\begin{equation}
\frac{t_s}{t_{\text{eq}}}= \frac{3\pi}{4}
\left(\frac{5\delta_{\text{eq}}}{3}\right)^{-3/2}, \quad
\frac{r_s}{r_{\text{eq}}}=\frac{3}{5\delta_{\text{eq}}},
\label{tsti}
\end{equation}
and the DM density perturbation at $t=t_{\rm eq}$ within the sphere containing a DM mass $M$ is $\delta_{\text{eq}}=2M_{\rm PBH}/M$. The time of the formal shell contraction into a point is $t_c\simeq2t_s$.

In reality, however, the DM shells do not remain spherically symmetric. There are directions of the radius vector in which the angular momentum of
the DM particles is zero~\cite{SikTkaWan97}. For example, if the tidal forces are created by the third PBH, then the angular momentum is zero in the direction of the radius vector of this PBH. The DM shell contracting along this direction will pass through the halo center (through the PBH pair). The nonspherical contraction of DM shells is analogous to the formation of “Zel’dovich pancakes” and the compression of baryonic matter into disks in disk galaxies along the rotation axis. However, in contrast to baryons, DM is not fixed in the disk, but passes through the halo center and experiences several damped oscillations (passages) before mixing with other shells. The above direction of zero angular momentum will change with time (it depends on the distribution of inhomogeneities generating the angular momentum), but at each instant of time, starting from $\sim t_{\rm eq}$, newer and newer DM shells will continuously pass through the center, creating a nonzero DM density at the center.

Consider a thin DM shell between stopping radii $r_s$ and $r_s+\delta r_s$ in the direction of zero angular momentum. The inner boundary of the shell will pass through the halo center at some time $t_c\simeq 2t_s$, while the outer boundary will pass with some time delay. Let us denote the thickness of this shell at the time of its passage through the center by $\delta r_c$. Accordingly, the contraction coefficient of the shell along its normal will be $\delta r_s/\delta r_c$. In contrast, in the transverse direction the shell can be represented for our estimate by a disk with a characteristic radius $r_t$, which we will find below. Thus, the DM density at the halo center is
\begin{equation}
\rho_c(t)=\rho_s(\tilde t)\frac{\delta r_s}{\delta r_c}\left(\frac{r_s(\tilde t)}{r_t}\right)^2,
\label{rholy}
\end{equation}
where $\tilde t\simeq t/2$, since the shell passing through the center at time $t$ under consideration stopped and began to contract at an earlier time $\tilde t$. The density $\rho_s(\tilde t)$ is a factor of $4$ higher than the density at the stopping radius at the time $t$ and a factor of $(3\pi/8)^2\approx1.388$ higher than the mean DM density in the Universe at
the time $\tilde t$.

The shell thickness at the passage time can be estimated as $\delta r_c\sim v_c\delta t_c$, where the passage velocity through the center is
\begin{equation}
v_c\simeq\left(\frac{2GM}{r_s}\right)^{1/2},
\label{vceq}
\end{equation}
where $M$ is the DM mass within the shell with the radius $r_s$. Thus,
\begin{equation}
\frac{\delta r_c}{\delta r_s}\simeq v_c\frac{\partial t_c}{\partial r_s}.
\end{equation}
Since $t_s\propto r_s^{9/8}$ and $t_c\approx2t_s$, $\partial t_c/\partial r_s=(9/4)t_s/r_s$. At $z\gg1$ (when the influence of dark energy is small) we may set $1+z_c\simeq2^{2/3}(1+z_s)$.

Note also that the mass within the shell $r_s$ is a factor of $(3\pi/4)^2=5.552$ higher than the DM mass in the shell of the same radius in a homogeneous Universe. Collecting all together, we find that the ratio
\begin{equation}
\frac{\delta r_c}{\delta r_s}=\frac{9\pi}{8}\approx3.5
\end{equation}
is constant and then
\begin{equation}
\rho_c(t)=\frac{\pi}{8}\bar\rho(\tilde t)\left(\frac{r_s(\tilde t)}{r_t}\right)^2,
\label{rhointer}
\end{equation}
where $\bar\rho(t)$ is the mean DM density in the Universe, and $\bar\rho(\tilde t)\simeq4\bar\rho(t)$ at $z\gg1$.

\section{INFLUENCE OF THE TIDAL FORCES PRODUCED BY INFLATIONARY PERTURBATIONS}

The quantity $r_t$ can contain the contribution from various tidal processes. In this section we will find the part of $r_t$ associated with the influence of inflationary DM density perturbations. A DM halo is formed around a PBH; the DM shells successively pass from expansion to contraction. However, because of the influence of tidal gravitational forces, the DM particles are displaced from the radial trajectory. Along the direction specified above, in which the accumulated angular momentum is zero, the shell passes through the center, while in the orthogonal direction the minimum distance of the particles from the PBH pair is equal to some characteristic value of $r_t$. Let us estimate this value. Consider a coordinate system with the center in the PBH pair. Following the method described in~\cite{BerDokEro03}, let us expand the gravitational potential
\begin{equation}
\phi(\vec r,t)= \phi_0\!+\!\left.\frac{\partial \phi}{\partial
r^i}\right|_0\!\!r^i
\!+\!\frac{1}{6}\left.\Phi_{ll}\right|_0\delta_{ij}r^ir^j
\!+\!\frac{1}{2}\left.T_{ij}\right|_0r^ir^j\!+\!\ldots\!, 
\label{series}
\end{equation}
where
\begin{equation}
\Phi_{ij}=\frac{\partial^2 \phi(\vec r)}{\partial r^i\partial r^j}, \quad
T_{ij}=\Phi_{ij}-\frac{1}{3}\Phi_{ll}\delta_{ij}.
\end{equation}
The tidal forces are created by the third term in~\eqref{series}.

It can be shown~\cite{BerDokEro03} that in the case of statistical averaging
\begin{equation}
\langle T_{ij}T_{ji}\rangle=\frac{2}{3}(4\pi)^2G^2{\bar\rho}^2\sigma^2(M), 
\label{ttsr}
\end{equation}
where $\sigma(M)$ is the rms value of the DM density perturbations in some scale of masses $M$ calculated using the power spectrum of the inflationary perturbations in DM. To calculate $\sigma(M)$, we use a power-law initial perturbation spectrum normalized to the Planck space telescope observational data.

We will use the method of successive approximations. In the first approximation the partial trajectory is radial; the particle moves only under the influence of the PBH pair and the mass of the spherically symmetric halo that grows around the pair. The particle velocity is
\begin{equation}
\frac{d\vec{v}_{\rm rad}}{dt}=-\nabla\phi (r),
\label{radeq}
\end{equation}
where $\phi(r)$ is the spherically symmetric part of the
potential. The solution of Eq.~\eqref{radeq} is given by Eq.~\eqref{param}.

In the second approximation the particle deviates from the radial trajectory, and its transverse displacement grows according to the equation
\begin{equation}
\frac{d^2r_{t,i}}{dt^2}=-T_{ij}(t)r^j,
\label{vshev1}
\end{equation}
where
\begin{equation}
T_{ij}(t)=T_{ij}(t_{\text{eq}})\left(\frac{t}{t_{\text{eq}}}\right)^{-4/3}.
\label{tlin}
\end{equation}
The particle initially recedes from the PBH and subsequently stops at $t=t_s$; the particle begins to move backward and approaches the PBHs to the minimum distance near $t=2t_s$. However, because of its transverse displacement, the particle does not fall into the BH, but passes at some minimum distance. Integrating Eq.~\eqref{vshev1}  until the time of its closest approach to the center, we get
\begin{equation}
r_{t,i}=-\int\limits_{t_{\rm eq}}^{2t_s}dt'\int\limits_{t_{\rm eq}}^{t'}dt''T_{ij}(t'')r^j(t'').
\end{equation}
In our numerical calculation, when integrating, it is convenient to pass from the time to a parameter $\theta$.

We fix the PBH mass $M_{\rm PBH}$. If we consider the cosmological redshift $z$, then, as follows from Eq.~\eqref{param}, the density perturbation at $t_{\rm eq}$
\begin{equation}
\delta_{\rm eq}=\frac{3}{5}\left(\frac{3\pi}{2}\right)^{2/3}\frac{1+z}{1+z_{\rm eq}}
\end{equation}
corresponds to the shell collapse at this time. The PBH creates such a perturbation in the DM mass
\begin{equation}
M=\frac{M_{\rm PBH}}{\delta_{\rm eq}}.
\end{equation}
This is the DM mass that lies within a sphere with a radius equal to the radial distance of the particle under consideration from the PBH. The initial radius of this shell at $t_{\rm eq}$ is
\begin{equation}
r_{\rm eq}=\left(\frac{3M}{4\pi \rho_{\rm eq}}\right)^{1/3}.
\end{equation}
For our estimate we will assume that the scale of the mass $M$ gives an inhomogeneity and non-sphericity creating the tidal forces and will consider the rms value of the perturbations. At $t_{\rm eq}$ this rms value is $\sigma_{\rm eq}(M)$. As a result, denoting the part of $r_t$ being considered in this section by $r_{t,in}$, we get
\begin{equation}
r_{t,in}=2^{9/4}3^{-1/4}\sigma_{\rm eq}(M)r_{\rm eq}\left(\frac{5\delta_{\text{eq}}}{3}
\right)^{-2}\varkappa(\delta_{\text{eq}}),
\label{rtinfl}
\end{equation}
where the function
\begin{eqnarray}
\varkappa(\delta_{\text{eq}})&=&\int\limits_{-\pi/2+(5\delta_{\rm eq}/3)^{1/2}}^{\pi/2}\!\!\!\!\!\!d\theta\cos^2\theta'\int\limits_{-\pi/2+(5\delta_{\rm eq}/3)^{1/2}}^{\theta'}\!\!\!\!\!\!\!\!\!\!d\theta\cos^4\theta
\nonumber
\\
&\times&\left(\frac{\pi}{2}+\theta+\frac{1}{2}\sin(2\theta)\right)^{-4/3}
\label{kapfun}
\end{eqnarray}
has an order $\simeq1$ and  $\varkappa(\delta_{\text{eq}}) \to 2.9$ when $\delta_{\text{eq}}\to0$. The ratio $r_{t,in}/r_s$ together with the contribution from the third PBH is shown in Fig.~\ref{gr1}.

\section{INFLUENCE OF THE THIRD PBH}

The angular momentum of the PBH pair is determined by the third nearest PBH. However, this third PBH influences the motion of the DM shells. Near $t_{\rm eq}$ the DM halo mass is still small and the inflationary perturbations did not grow. Therefore, one might expect that at larger $f$ the third PBH will make a relatively large contribution. Let us find $r_t$ that arises due to the tidal effect of the third PBH. Since the distribution of distances to the third PBH is approximately flat, the characteristic comoving distance to the third PBH is $y_{ch}\sim\bar x/2$, where the mean separation between the PBHs is $y_{ch}\sim\bar x/2$, where the mean separation between the PBHs is
\begin{equation}
\bar{x}\simeq\left(\frac{3M_{\rm PBH}}{4\pi f\rho_{\rm eq}}\right)^
{1/3}.
\end{equation}

The evolution of $r_t$ under the influence of the third PBH is defined by the equation (the equation for the tidal force)
\begin{equation}
\frac{d^2r_t}{dt^2}=-\frac{2GM_H(t)}{(y_{ch}s(t))^3}r(t),
\label{vshev}
\end{equation}
where $r(t)$, as above, is determined by the first approximation (strictly radial motion), and the mass of the DM halo around the third PBH grows as
\begin{equation} 
M_H(z)=\frac{3}{2}\left(\frac{2}{3\pi}\right)^{2/3} \frac{1+z_{\rm
eq}}{1+z}M_{\rm PBH}.
\label{rcol2}
\end{equation}
Integrating Eq.~\eqref{vshev}, twice, we obtain $r_t$ as a function of $f$:
\begin{equation}
r_{t,3}=\frac{12}{\pi^{2/3}}t_{\rm eq}^2r_{\rm eq}\left(\frac{5\delta_{\text{eq}}}{3}
\right)^{-2}\frac{GM_{\rm PBH}}{y_{ch}^3}\varkappa(\delta_{\text{eq}}).
\label{rt3pbh}
\end{equation}

As the DM particles approach the PBH pair, their scattering by the gravitational field of the pair begins to dominate. Therefore, the radius $r_t$ is limited from below by the “radius of influence” of the PBH pair:
\begin{equation} 
r_{\rm infl}\simeq \frac{2GM_{\rm PBH}}{v_c^2}\simeq r_s(z)\frac{M_{\rm PBH}}{M(z)}.
\end{equation}

\begin{figure}[H]
\centering
\includegraphics[angle=0,width=0.49\textwidth]{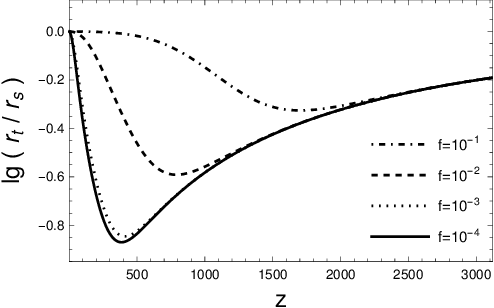}
\caption{The characteristic DM shell radius $r_t$ determined by the combined effect of the tidal forces from the inflationary perturbations and the third nearest PBH with respect to the shell stopping radius $r_s$ at redshift $z$. The curves correspond to the PBH mass $M_{\rm PBH}=30M_\odot$ and (from top to bottom) $f=10^{-1}$, $f=10^{-2}$, $f=10^{-3}$, and $f=10^{-4}$.}
\label{gr1}
\end{figure}

Since the inflationary perturbations, the third PBH, and the radius of influence act independently and since $r_t$ cannot exceed $r_s$, in our final calculation of the density it is necessary to set
\begin{equation} 
r_t={\rm min}\left\{r_s, \sqrt{r_{t,in}^2+r_{t,3}^2+r_{\rm infl}^2}\right\}.
\end{equation}
This scale is shown in Fig.~\ref{gr1}, where the smoothing $1/r_t^2=1/r_s^2+1/(r_{t,in}^2+r_{t,3}^2+r_{\rm infl}^2)$ was done. This smoothing is not fundamentally important for our estimation, since at $r_t\sim r_s$ the accuracy of the above expression in any case is low, but the smoothing is useful in the subsequent numerical integration of the equations of motion for the pair.

As can be seen from Fig.~\ref{gr1}, at $f\leq10^{-3}$ the dependence on $f$ is weak. This is because $\bar x$ increases as $f$ decreases, the third PBH is farther and farther from the PBH pair under consideration, and its gravitational influence becomes considerably weaker than the influence of the inflationary perturbations. A boundary value of $f\sim10^{-3}$ could be expected, since $f$ characterizes the relative density perturbation $\sim M_{\rm PBH}/M$ created by the PBH near $t_{\rm eq}$, while the value of the inflationary perturbations on the scales under consideration at $t_{\rm eq}$ is exactly $\sim10^{-3}$.

\begin{figure}[H]
\centering
\includegraphics[angle=0,width=0.49\textwidth]{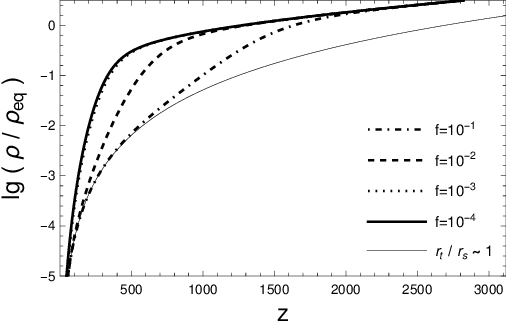}
\caption{The density at the center of the DM halo surrounding a PBH pair. The designations are the same as those in Fig.~\ref{gr1}. The lower solid curve indicates the density $\rho_c(t)=(\pi/2)\bar\rho(t)$.}
\label{gr2}
\end{figure}

Substituting $r_t$ into Eq.~\eqref{rhointer}, we find the DM density
at the halo center at each instant of time (see Fig.~\ref{gr2}).
This DM at the center will affect the PBH pair, taking
away energy from it.

\section{EVOLUTION OF THE PBH PAIR ORBIT}

Using a numerical experiment, \cite{Quinlan:1996vp} obtained the law of evolution of the semimajor axis of a black hole pair under the influence of incoming stars with the same velocity and then performed an averaging over the velocities under the assumption of their Maxwellian distribution. In our case, it is necessary to use the expression for the evolution of the semimajor axis before the averaging, since we consider the passage of the DM shells through the center with one velocity $v_c$. Thus, in the case being considered by us, the expression from~\cite{Quinlan:1996vp} is
\begin{eqnarray} 
\frac{d}{dt}\left(\frac{1}{a}\right)=H_1\frac{G\rho}{v_c},
\label{eq:a_dec1}
\end{eqnarray} 
where $H_1\simeq20$, and $v_c$ is calculated from Eq.~\eqref{vceq}. The rate of change in the orbital eccentricity of the PBH pair is \cite{Quinlan:1996vp}
\begin{equation}
    \frac{d e}{ d \ln{(1/a)}} = K_1,
\end{equation}
where $K_1$ depends on $e$ (in particular, it contains a positive power of $1-e^2$ as a factor), and the fitting formula for $K_1$ is given in~\cite{Quinlan:1996vp}. In the case being considered by us, $1-e^2\ll 1$. Therefore, the change in eccentricity under the influence of the passage of DM shells may be neglected with a good accuracy.

In our numerical calculation we divide the evolution process into two parts: the initial stage, when the pair energy loss according to~\eqref{eq:a_dec1}  dominates, and the subsequent stage of orbital decay under the action of the losses through gravitational radiation. At the first stage Eq.~\eqref{eq:a_dec1} is integrated numerically using the laws of evolution of $\rho$ and $v_c$ with time found in the preceding sections. To consider the second stage, we use the well-known expression for the pair contraction time under the influence of gravitational radiation
\begin{equation}
t_c=\frac{3c^5}{170G^3M_{\rm BH}^3}a^4(1-e^2)^{7/2}.
\label{tc}
\end{equation}
The quantities found at the end of the first stage of evolution are substituted into this expression as $a$  and $e$. Subsequently, we compare the sum (\ref{tc}) and the durations of the first stage with the current time $t_0$ and find the time derivative of the merger probability $dP(t_0)/dt_0$, which is used in the next section to calculate the rate of gravitational bursts.

\section{THE PBH MERGER RATE}

\begin{figure}[H]
\centering
\includegraphics[angle=0,width=0.49\textwidth]{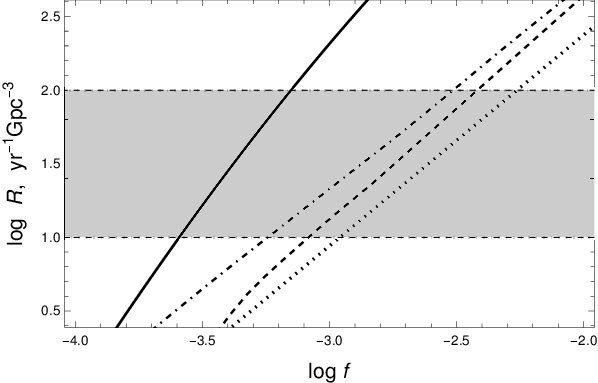}
\caption{The rate of gravitational bursts from the mergers of PBHs in pairs. The dotted and solid curves indicate the rate without and with the influence of the DM halo on the PBH pair, respectively. The gray color marks the region admissible by the observations of gravitational bursts. The dashed curve indicates the merger rate calculated by~\cite{PilTkaIva22}. The dash–dotted curve was obtained in our conservative estimation.}
\label{gr3}
\end{figure}

Following the approach of~\cite{Naketal97,Ioketal98,Sasetal16}, let us write out the basic relations required to calculate the statistics of PBH mergers in pairs by taking into account the influence of the DM halo. Denote the comoving separations between the components of the PBH pair and between the center of mass of the pair and the third PBH by $x$ and $y$, respectively. The PBH pair is formed at the radiation-dominated stage at $t<t_{\rm eq}$, and we assume that the scale factor of the Universe $s(t)$ is normalized as $s(t_{\rm eq})=1$. The condition for the formation of a gravitationally bound pair is~\cite{Naketal97,Sasetal16}
\begin{equation}
\frac{M_{\rm PBH}}{x^3s_m^3}\sim\rho_r,
\end{equation}
where $s_m=(1/f)(x/\bar{x})^3$ and $\rho_r$ is the radiation density. According to the model~\cite{Naketal97,Ioketal98,Sasetal16}, the semiminor axis of the orbit is determined by the tidal effect of the third PBH (the contribution of the inflationary perturbations is minor~\cite{Ero16-1}). As a result, the semimajor and semiminor axes of the pair are~\cite{Naketal97,Ioketal98,Sasetal16}
\begin{equation}
a=\alpha\frac{1}{f}\frac{x^4}{\bar{x}^3},\qquad b=\beta a\left(\frac{x}{y}\right)^3,
\label{ab}
\end{equation}
where $\alpha\sim1$ and $\beta\sim1$. The eccentricity of the orbit is $e=(1 - \beta^2x^6/y^6)^{1/2}$. The quantities $x$ and $y$ have a flat probability distribution
\begin{equation}
dP=\frac{18x^2y^2}{\bar{x}^6}dxdy.
\label{dpdxdydnu}
\end{equation}
Using this distribution, we can find the pair merger probability in a time $<t$:
\begin{equation}
P(<t)=\frac{\alpha}{\beta}\left[\frac{37}{29}\left(\frac{t}{t_{\rm max}}\right)^{3/37}-\frac{8}{29}\left(\frac{t}{t_{\rm max}}\right)^{3/8}\right],
\label{pint}
\end{equation}
where
\begin{equation}
t_{\rm max}=\frac{5c^5}{512G^3M_{\rm PBH}^3}\frac{\alpha^4}{\beta^{16/3}}\frac{\bar x^4}{f^4}.
\label{tmax}
\end{equation}
As a result, we calculate the pair merger rate
\begin{equation}
R=\left.\frac{\rho_c\Omega_mf}{M_{\rm PBH}}\frac{dP(<t)}{dt}\right|_{t=t_0},
\label{rate1}
\end{equation}
where $\rho_c=9.3\times10^{-30}$~g~cm$^{-3}$ is the critical density, and $\Omega_m\approx0.27$  is the cosmological DM density parameter. The merger rate for the solution~\eqref{pint} is indicated in Fig.~\ref{gr3} by the dotted curve.

In our calculation, instead of~\eqref{pint} we numerically calculate the probability based on~\eqref{dpdxdydnu}. On the plane of parameters $x$ and $y$ we find the area of the region in which the pair merger time is less than the current time $t_0$ by taking into account the two stages of evolution of the pair orbit described in the previous section. The result of our calculation presented in Fig.~\ref{gr3} shows that nonspherical halo contraction enhances the merger rate by several times compared to the calculations in which this effect is ignored.

\section{A CONSERVATIVE ESTIMATE OF THE MERGER RATE}

In this section, for comparison, we will give a more conservative estimate of the merger rate without including the one-dimensional DM shell contraction, but by taking into account the angular momentum distribution of DM particles beginning from zero. Particles with a low angular momentum can pass through the orbit of the PBH pair, causing the energy of its orbital motion to be lost.

The DM shell with a mass $dM$ decoupling from the cosmological expansion contains particles with various angular momenta, including zero one. DM particles with an angular momentum $l^2 < l^2_{\rm crit} = 4 a G M_{\rm PBH}$ are located in the loss cone and fly inside the orbit of the binary PBH. These particles after their scattering by the binary extract an energy $\Delta E = \eta G M_{\rm PBH} dM f_{\rm lc} /a$ from it, where $\eta \sim 1$ and $f_{\rm lc}$ shows what fraction of DM particles is located inside the loss cone. The contraction rate of the semimajor axis is then defined by the equation
\begin{equation}
    \dot{a}_{\rm lc} = - a f_{\rm lc}\frac{\dot{M_{\rm H}}}{M_{\rm PBH}},
\end{equation}
where $\dot{M_{\rm H}}$ is the growth rate of the DM halo mass obtained by differentiating Eq.~\eqref{rcol2} with respect to time. For the same reasons as above, we neglect the change in eccentricity because of the interaction with DM particles.

Let us now determine $f_{\rm lc}$. We will assume that the angular momentum distribution has an almost isotropic form $dp \propto l dl$, but is truncated from above by some value of $l_{\rm max}$ (as in~\cite{SikTkaWan97}), which in our case is determined by the tidal effects from the inflationary perturbations and the third PBH. Since the greatest increase in the angular momentum of a DM particle occurs near the stopping time of DM shell expansion, when the shell has a maximum radius, the angular momentum created by the inflationary perturbations can be estimated as
\begin{equation}
    l_{\rm t, in} = \sqrt{\langle T_{ij}T_{ji}\rangle} r^2_s t.
\end{equation}
The angular momentum from the third PBH will be
\begin{equation}
    l_{\rm t,3} \sim r_s F_t t = \frac{2 r_s^2 G M_{\rm H}}{(y s(t))^3} t,
\end{equation}
where $F_{t}$  is the tidal force per unit mass from the third PBH \eqref{vshev}, and $y$ is the distance to the third PBH at $t_{\rm eq}$. Recall that this distance fixes the eccentricity of the binary system. Thus, the total maximum angular momentum will be
\begin{equation}
    l_{\rm max} = \sqrt{l^2_{\rm t,3} + l^2_{\rm t, in}}.
\end{equation}
Note that the angular momentum of the DM particles cannot be greater than the angular momentum in a circular orbit of radius $r_s$, i.e., $l^2_{\rm circ} = G M_{\rm H} r_s$. Therefore, the natural constraint $l_{\rm max} < l_{\rm circ}$ should be taken into account. However, the latter is true only at relatively low redshifts, where the contribution of the DM particles to the orbit evolution is negligible.

To calculate the merger rate, it is necessary to find the lifetime of the binary system. For this purpose, we solve the following system of equations:
\begin{equation}
    \dot{a} = \dot{a}_{\rm gw} + \dot{a}_{\rm lc},
    \label{a_decay}, 
\end{equation}   
\begin{equation}
    \dot{e} =\dot{e}_{\rm gw}, \label{e_decay}
\end{equation}
where the rates of change in the semimajor axis and eccentricity of the binary system during gravitational wave radiation are described by the equations
\begin{equation}
    \dot{a}_{\rm gw} = - \frac{128}{5} \frac{G^3 M^3_{\rm BH}}{c^5 a^3} \frac{1}{(1 - e^2)^{7/2}} \left ( 1 + \frac{73}{24} e^2 + \frac{37}{96} e^4 \right),
\end{equation}    
\begin{equation}    
    \dot{e}_{\rm gw} = -\frac{608}{15} \frac{G^3 M^3_{\rm BH}}{c^5 a^4} \frac{e}{(1 - e^2)^{5/2}} \left ( 1 + \frac{121}{304}e^2 \right).
\end{equation}

Subsequently, we numerically find the region in the space of initial $a$ and $e$ of the binary systems that merge by the time $t$, calculate the double integral of Eq.~\eqref{dpdxdydnu} over this region. And then differentiating the resulting expression with respect to time, we find the PBH merger rate. The result of our calculation is indicated in Fig.~\ref{gr3} by the dash-dotted line. 

\section{CONCLUSIONS}

In this paper we considered the influence of the DM flow continuously passing through the center of a DM halo on the evolution of a PBH pair. The DM particles in the halo forming around the PBH pair acquire an angular momentum under the influence of tidal gravitational forces. Neighboring PBHs, primarily the third nearest PBH, and inflationary density perturbations are the source of these forces. If the angular momentum is nonzero, then a DM particle will pass not through the halo center, but at some minimum distance. However, in the angular momentum distribution there is always a direction with zero angular momentum~\cite{SikTkaWan97}. In this direction the DM shell passes through the center, although it can deform to some extent. After several oscillations, the shell is dispersed and becomes part of the spherically symmetric DM distribution in the halo. We calculate the characteristic contraction of the shell during its passage through the center in comparison with the thickness of the same shell at the stopping time of its expansion. We also calculate the characteristic distance of the particle passage by the halo center. This allows the density of the DM flow continuously passing through the halo during its formation to be estimated. The flow ceases only when the halo growth and the detachment of the new DM shells from the cosmological expansion cease.

Our calculations confirm the conclusion drawn by~\cite{PilTkaIva22} through numerical simulations that the DM halo accelerated the mergers of PBH pairs. In contrast to~\cite{PilTkaIva22}, in our calculations we took into account the passages of background (not highly clustered) DM through the halo center and showed that this DM component enhances the influence of the halo by several more times. As a result, when only these effects are taken into account, a fraction of PBHs in DM $f\sim(3\div7)\times10^{-4}$ is enough to explain the LIGO/Virgo/KAGRA observational data. Note, however, that the total rate of gravitational wave bursts can be affected by additional factors, for example, those associated with PBH clustering~\cite{StaBel23,Sta24} and leading to a increase in the admissible values of $f$. All of the effects can be taken into account in future in numerical simulations with a high mass resolution that can describe the DM inhomogeneities both on the scale of large DM halos containing many PBHs and on the scales that correspond to the density gradient in the DM shells passing through the center of small halos around PBH pairs.

This study was supported by Russian Science Foundation grant no. 23-22-00013, https://rscf.ru/project/23-22-00013/.


\begin{thebibliography}{99}

\bibitem{ZelNov67}  Zel’dovich Ya.B. and  Novikov I.D., Sov. Astron. {\bf 10}, 602 (1967).

\bibitem{Haw71}  Hawking S., MNRAS {\bf 15}, 75 (1971).

\bibitem{KhlPol0}  Khlopov M.Yu., Polnarev A.G., Phys. Lett. {\bf B 97}, 383 (1980).

\bibitem{BerKuzTka83}  Berezin V.A., Kuzmin V.A., Tkachev I.I., Phys. Lett. B {\bf 120}, 91 (1983).

\bibitem{Jed97}  Jedamzik K., Phys. Rev. D {\bf 55}, R5871(R) (1997).

\bibitem{DolSil93}  Dolgov A.,  Silk J., Phys. Rev. D {\bf 47}, 4244 (1993).

\bibitem{RubKhlSak00}  Rubin S.G.,  Khlopov M.Yu.,  Sakharov A.S., Grav. Cosmol. S {\bf 6}, 51 (2000).

\bibitem{RubSakKhl01}  Rubin S.G.,  Sakharov A.S.,  Khlopov M.Yu., JETP {\bf 92}, 921 (2001).

\bibitem{Naketal97}  Nakamura T.,  Sasaki M.,  Tanaka T.,  Thorne K.S., The Astrophys. J. {\bf 487}, L139 (1997).

\bibitem{Ioketal98}  Ioka K.,  Chiba T.,  Tanaka T.,  Nakamura T., Phys. Rev. D {\bf 58},  063003 (1998).

\bibitem{Sasetal16}  Sasaki M.,  Suyama T.,  Tanaka T., Yokoyama S., Phys. Rev. Lett.  {\bf 117}, 061101 (2016).

\bibitem{Doletal20}    Dolgov A. D. et al., Journal of Cosmology and Astroparticle Physics {\bf 12}, 017 (2020).

\bibitem{CarKuh20}  Carr B., Kuhnel F., Annual Review of Nuclear and Particle Science {\bf 70}, 355 (2020).

\bibitem{Ero16-1}  Eroshenko Yu.N., J. Phys.: Conf. Ser. {\bf 1051}, 012010 (2018).

\bibitem{Ero16-2}  Eroshenko Yu.N., Astronomy Letters {\bf 42}, 347 (2016).

\bibitem{DokEro01} Dokuchaev V.I., Eroshenko Y.N., Astronomy Letters {\bf 27}, 759 (2001).

\bibitem{DokEro03} Dokuchaev V.I.,  Eroshenko Y.N., Astronomical \& Astrophysical Transactions {\bf 22}, 727
(2003).

\bibitem{Hayetal09} Hayasaki K., Takahashi K., Sendouda Y., Nagataki S., Publications of the Astronomical Society of Japan {\bf 68}, 66 (2016).

\bibitem{StaBel23} Stasenko V., Belotsky K., MNRAS {\bf 526}, 4308 (2023).

\bibitem{Sta24} Stasenko V., Phys. Rev. D {\bf 109}, 123546 (2024).

\bibitem{SikTkaWan97} Sikivie P., Tkachev I.I., Wang Y., Phys. Rev. D {\bf 56}, 1863 (1997).

\bibitem{TkaPilYep20} Tkachev M.V.,  Pilipenko S.V., Yepes G., Mon. Not. R. Astron. Soc. {\bf 499}, 4854 (2020).

\bibitem{PilTkaIva22} Pilipenko S.,  Tkachev M.,  Ivanov P., Phys. Rev. D {\bf 105}, 123504 (2022).

\bibitem{BerDokEro03} Berezinsky V.,  Dokuchaev V., Eroshenko Y., Phys. Rev. D {\bf 68}, 103003 (2003).

\bibitem{Quinlan:1996vp} Quinlan G.D., New Astron. {\bf 1}, 35 (1996).

\end{thebibliography}
\end{document}